\documentclass[reprint,groupedaddress,aps,prl,superscriptaddress]{revtex4-2}
\usepackage{graphicx}
\usepackage{xcolor}
\usepackage{dcolumn}
\usepackage{rotating}
\usepackage{makecell}
\usepackage{enumerate} 
\usepackage{bm}
\usepackage{bbm}
\usepackage{amsmath}
\usepackage{amsfonts}
\usepackage{tikz}
\usepackage[version=4]{mhchem}
\usetikzlibrary{quantikz}
\usepackage[justification=Justified]{caption}
\usepackage[justification=Justified]{subcaption}
\usepackage{ragged2e}
\usepackage{booktabs}
\usepackage{siunitx}
\usepackage{multirow}
\usepackage{tabularx}
\usepackage{tikz}
\usepackage{circuitikz}
\usepackage{mhchem}
\usetikzlibrary{patterns}
\usetikzlibrary{decorations.markings,  decorations.pathmorphing, decorations.pathreplacing, decorations.shapes, arrows.meta}
\usepackage{minitoc}

\bibpunct{}{}{,}{s}{}{\textsuperscript{,}}

\newcommand{\ii}{\mathbbm{i}}

\renewcommand{\Re}{\mathrm{Re}}

\newcommand{\br}{\bm{r}}
\newcommand{\bx}{\bm{x}}
\newcommand{\bn}{\bm{n}}
\newcommand{\lan}{\langle}
\newcommand{\ran}{\rangle}
\newcommand{\nl}{\nonumber\\}

\newcommand{\bU}{\mathbf{U}}

\newcommand{\bUbar}{\mathbf{\overline{U}}}
\newcommand{\nbar}{\overline{n}}
\newcommand{\bnbar}{\bm{\overline{n}}}
\newcommand{\bs}{\mathbf{s}}
\newcommand{\hS}{\hat{S}}

\begin{document}


\title{Spin-adapted neural-network backflow for symmetry-preserving simulations of strongly correlated electrons}
\author{Yunzhi Li}
\author{Zibo Wu}
\author{Bohan Zhang}
\author{Wei-Hai Fang}
\author{Zhendong Li}\email{zhendongli@bnu.edu.cn}
\affiliation{Key Laboratory of Theoretical and Computational Photochemistry, Ministry of Education, College of Chemistry, Beijing Normal University, Beijing, 100875, China}
\affiliation{Institute for Advanced Study, Beijing Normal University, Beijing, 100875, China}

\begin{abstract} 
Strongly correlated molecules often contain dense manifolds of low-lying spin states, making total-spin symmetry essential for predictive electronic-structure theory. Neural-network quantum states provide flexible variational wavefunctions, but commonly used fermionic architectures do not enforce this symmetry and can therefore converge to spin-contaminated states with misleading energies and properties. Here we introduce a spin-adapted neural-network backflow (SA-NNBF) ansatz in second quantization, which combines configuration-dependent spatial orbitals with a compressed spin eigenfunction. A projected tensor compression scheme for spin eigenfunctions and a particle-hole representation make variational Monte Carlo calculations with SA-NNBF practical for active spaces containing more than one hundred electrons. Across hydrogen chains and iron–sulfur clusters, SA-NNBF eliminates spin contamination and consistently achieves lower variational energies than standard NNBF with a comparable number of parameters. For the CAS(113e,76o) active-space model of FeMoco, SA-NNBF yields a highly compact spin-adapted variational state, achieving an energy competitive with recent spin-adapted DMRG calculations at bond dimension $D=10000$ while using orders of magnitude fewer parameters. Our work establishes a general framework for developing spin-symmetry-preserving neural-network quantum states for chemically realistic strongly correlated electrons.
\end{abstract}

\maketitle

\let\oldaddcontentsline\addcontentsline
\renewcommand{\addcontentsline}[3]{}

\section{Introduction}

Strongly correlated electrons are responsible for the low-energy physics of many catalytic and magnetic molecules, but they remain difficult to describe because several electronic configurations and spin states can become nearly degenerate. In such systems, total-spin symmetry is more than a formal constraint. It determines the identity of the targeted electronic state and controls spin-dependent observables, and even a small symmetry violation can mix different spin states that should be orthogonal. This issue is particularly acute for polynuclear transition-metal complexes, where different spin states are often separated by only a few milli-Hartree\cite{sharma2014low,Li2017LMO,li2019electronic2}.

Active-space electronic-structure methods provide a systematic framework for treating such near-degeneracies by explicitly correlating the frontier orbitals involved in spin coupling and bond rearrangement. Among these methods, density matrix renormalization group (DMRG) algorithms, particularly spin-adapted implementations, have become a powerful tool for large active spaces\cite{sharma2012spin}. Nevertheless, strongly correlated systems such as the FeMo-cofactor (FeMoco) can still exhibit substantial entanglement, requiring very large bond dimensions for quantitative accuracy\cite{Li2025EMO,zhai2026classical}. This motivates the development of alternative wavefunction parameterizations that preserve spin symmetry while offering a more compact representation.

Neural-network quantum states (NQS) provide such an alternative by representing many-electron amplitudes with expressive neural architectures\cite{Carleo2017,choo2020fermionic,Hermann2023_reviews}. Fermionic neural-network backflow (NNBF) ans\"atze are among the most accurate second-quantized NQS forms for ab initio quantum chemistry\cite{luo2019backflow,Liu2024,liu2024unifying,Liu2025,Shang2025transformer,
gu2025solving,rende2026transformer}, which generalizes Slater determinants by introducing configuration-dependent spin-orbitals generated by neural networks. 
Yet their flexibility also creates a fundamental vulnerability: the standard construction does not guarantee that the optimized state is an eigenfunction of $\hat{S}^2$. In near-degenerate spin manifolds, variational optimization can then lower the energy by mixing undesired spin sectors, leading to quantitatively inaccurate energies and qualitatively wrong spin correlations.
 
Despite the central role of spin symmetry in strongly correlated open-shell systems, incorporating total-spin symmetry into neural-network wavefunctions remains challenging. Only a few attempts have been reported so far\cite{Li2024SpinSymmetryEnforced,li2025spin,vieijra2020Su2RBM}.
For molecular electronic structure, recent real-space neural wavefunctions have enforced spin symmetry through modified optimization objectives\cite{Li2024SpinSymmetryEnforced} or first-quantized spin-adapted antisymmetrization\cite{li2025spin}. 
These advances establish the importance of spin symmetry, but they do not provide a general ansatz-level construction for second-quantized molecular Hamiltonians. 
Vieijra et al.\cite{vieijra2020Su2RBM} imposed SU(2) symmetry in a restricted Boltzmann machine (RBM) for the one-dimensional antiferromagnetic Heisenberg model
by working in a spin-coupled basis. However, the spin-coupled representation generally leads to non-sparse Hamiltonians for molecular electronic structure, and is therefore not well suited for variational Monte Carlo\cite{becca2017quantum} (VMC). These limitations motivate the development of a symmetry-preserving neural-network ansatz in second quantization that retains the expressive power of modern NQS architectures while enforcing total-spin symmetry.
 
Here we develop a spin-adapted neural-network backflow (SA-NNBF) framework for strongly correlated electrons in second quantization. The ansatz separates the configuration-dependent spatial orbitals generated by a neural network from an explicitly constructed spin eigenfunction, and then recombines them into a fully antisymmetric many-electron wavefunction. It can be viewed as a second-quantized generalization of the spin-adapted antisymmetrization method recently introduced in first quantization\cite{li2025spin}, tailored to active-space electronic-structure problems. To make the total spin constraint computationally practical, we introduce two algorithmic ingredients. First, we develop a projected tensor compression procedure that represents spin eigenfunctions as a short sum of product states, greatly reducing the number of terms required relative to the exact spin decomposition. Second, by exploiting particle-hole duality in second quantization\cite{li2016hilbert}, we introduce a compact representation for more-than-half-filled active spaces, which reduces redundant variational parameters and eases optimization without sacrificing expressivity. Together with an efficient semi-stochastic local-energy evaluation algorithm\cite{wu2025hybrid}, these ingredients enable VMC calculations with SA-NNBF for molecular active spaces containing more than one hundred electrons. 
 
We benchmark SA-NNBF on prototypical strongly correlated systems, including hydrogen chains\cite{hachmann2006multireference} and iron-sulfur clusters\cite{sharma2014low,Li2017LMO,li2019electronic2}, and then apply it to the CAS(113e,76o) active-space model of FeMoco\cite{li2019electronic2}. Across these systems, enforcing spin symmetry removes the spin contamination observed in standard NNBF and consistently improves the reliability of both variational energies and spin-related observables. For FeMoco, SA-NNBF yields a highly compact spin-adapted variational state with substantially fewer parameters, while achieving a lower energy than recent large-bond-dimension spin-adapted DMRG (SA-DMRG) calculations at bond dimension $D=10000$\cite{Li2025EMO}. These results demonstrate that spin adaptation can turn neural-network quantum states into compact, accurate, and symmetry-preserving tools for simulating chemically realistic strongly correlated systems.

\section{Results}

\subsection{Spin-adapted neural-network backflow ansatz}
We consider an $N$-electron molecular system described by a spin-restricted basis containing $2K$ spin-orbitals $\{\chi_k\}$ ($k=1,2,...,2K$), where
\begin{align}
  \chi_k(\bx) = 
  \begin{cases}
    \overline{\chi}_i(\br) \alpha(\sigma), \quad k = 2i - 1, \\
    \overline{\chi}_i(\br) \beta(\sigma), \quad k = 2i. \\
  \end{cases} \label{eq: restricted-basis}
\end{align}
Here, $\bx\equiv(\br, \sigma)$ is the spatial-spin full coordinate of a single electron; $\{\overline{\chi}_i(\br)\}$ are spatial basis functions; $\alpha$ and $\beta$ are the spin-up and spin-down eigenfunction for the spin of a single electron.
Therefore, the $N$-electron wavefunction in second quantization $\Psi(\bn)$ can be written as a function of the occupation number vector $\bn \equiv (n_1, n_2, ..., n_{2K})$, a binary vector where $n_k$ is $0$ (1) for the empty (occupied) $k$-th spin-orbital and $\sum_{k=1}^{2K} n_k = N$.

\begin{figure*}[t]
    \centering
    \includegraphics[width=1.0\linewidth]{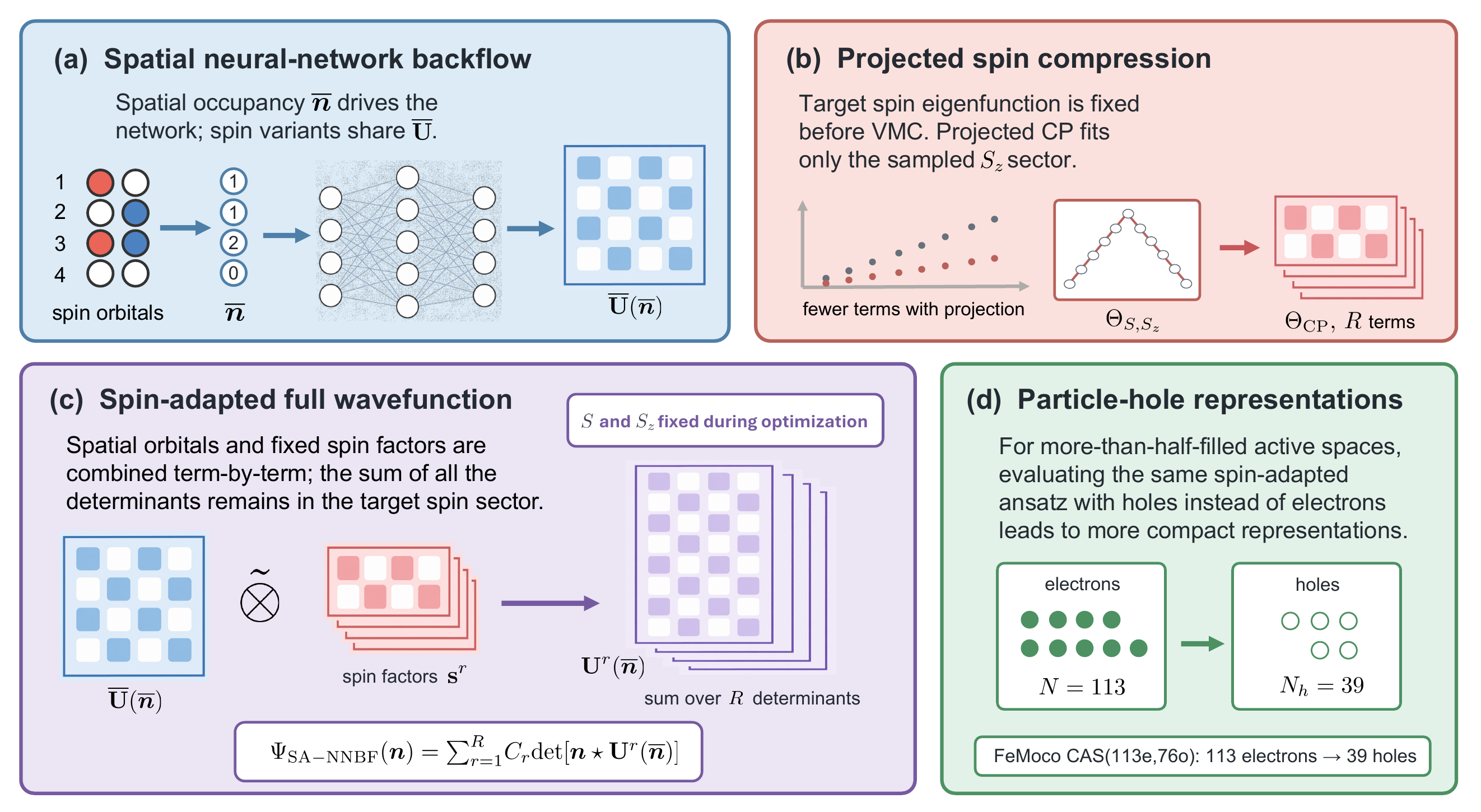}
    \caption{Overview of the SA-NNBF ansatz and the two reductions that make it scalable. (a) The neural network takes the spatial occupation vector $\bnbar$ as input and outputs configuration-dependent spatial orbitals $\bUbar(\bnbar)$, so configurations with different spin assignments but the same spatial occupancy share the same spatial orbitals. (b) The target spin eigenfunction is compressed into a projected CP sum of products in the fixed-$S_z$ sector before VMC optimization. (c) The spatial orbitals and fixed spin factors are combined term by term to form the final wavefunction $\Psi_{\rm SA\text{-}NNBF}(\bn)$. (d) For more-than-half-filled active spaces, the same construction is evaluated in the hole representation; in FeMoco with the CAS(113e,76o) active space\cite{li2019electronic2}, this reduces the determinant dimension from 113 electrons to 39 holes.}
    \label{fig: scheme}
\end{figure*}

The SA-NNBF architecture for approximating $\Psi(\bn)$ is illustrated in Fig. \ref{fig: scheme}. Standard NNBF generates configuration-dependent spin-orbitals\cite{luo2019backflow}. SA-NNBF instead generates configuration-dependent spatial orbitals, collected in a $K \times N$ matrix $\bUbar(\bnbar)$, from the spatial occupation vector $\bnbar \equiv(\nbar_1,\nbar_2,\ldots,\nbar_K)$ with $\nbar_j=n_{2j-1}+n_{2j}$. Configurations that differ only by spin assignments therefore share the same spatial orbitals, which is the key structural constraint that allows the neural-network component to be combined with a fixed spin eigenfunction. In the present implementation, $\bUbar(\bnbar)$ is generated by an embedding layer followed by a one-hidden-layer feed-forward neural network; more expressive architectures can be substituted without changing the spin-adapted construction.

The spin component is chosen as a target spin eigenfunction $\Theta$ with total spin $S$ and spin projection $S_z$. Any such $\Theta$ can be written, exactly or to controlled numerical accuracy, in the sum-of-products form
\begin{align}
  \Theta(\sigma_1, \sigma_2, ..., \sigma_N) = \sum_{r=1}^{R} C_r 
  \prod_{j=1}^N \theta^r_{j}(\sigma_j),\label{eq: Theta-SAAM}
\end{align}
where each one-electron spinor is parameterized as
\begin{align*}
  \theta^{r}_{j}(\sigma) = s^r_{1j} \alpha(\sigma) + s^r_{2j} \beta(\sigma).
\end{align*}
The coefficients $C_r$ and matrices $\bs^r$ are fixed before VMC optimization; only the spatial backflow orbitals are variational. This separation prevents the neural network from changing the target spin sector during energy optimization.

With the spatial part and the spin part specified, $R$ spin-orbital coefficient matrices $\bU^r(\bnbar)$ are formed by 
\begin{align}
  \bU^r(\bnbar) = & \bUbar(\bnbar) \, \tilde{\otimes} \, \bs^r,
\end{align} 
where the operator $\tilde{\otimes}$ is defined as
\begin{align}
  (\mathbf{\overline{U}}(\bnbar) \, \tilde{\otimes} \, \mathbf{s}^r)_{kj} \equiv 
  \begin{cases}
    \overline{U}_{ij}(\bnbar)\cdot s^r_{1j}, \quad k = 2i-1, \\
    \overline{U}_{ij}(\bnbar)\cdot s^r_{2j}, \quad k = 2i. \\
  \end{cases} \label{eq:tildeProduct}
\end{align}
The resulting SA-NNBF amplitude is
\begin{align}
  \Psi_{\rm SA\text{-}NNBF}(\bn) = \sum_{r=1}^{R} C_r \cdot {\rm det}[\bn \star \bU^r(\bnbar)], \label{eq: Psi-SA-NNBF-red}
\end{align}
where $\star$ selects the rows associated with occupied spin-orbitals in $\bn$. 
If the sum-of-products representation of $\Theta$ is exact, Eq. \eqref{eq: Psi-SA-NNBF-red} is also an eigenfunction of both $\hat{S}^2$ and $\hat{S}_z$ for any choice of the spatial orbitals $\bUbar(\bnbar)$. In practical calculations we use the compressed spin function described below.
Finally, we note that although this work uses a single set of spatial orbitals $\bar{\mathbf{U}}(\bm{n})$ for each $\bm{n}$, the ansatz can in principle be extended to include multiple sets of spatial orbitals to further enhance its expressibility, as is commonly done in NNBF\cite{luo2019backflow}.


\subsection{Projected tensor compression for spin eigenfunctions}

\begin{figure}[t]
    \centering
    \includegraphics[width=1.0\linewidth]{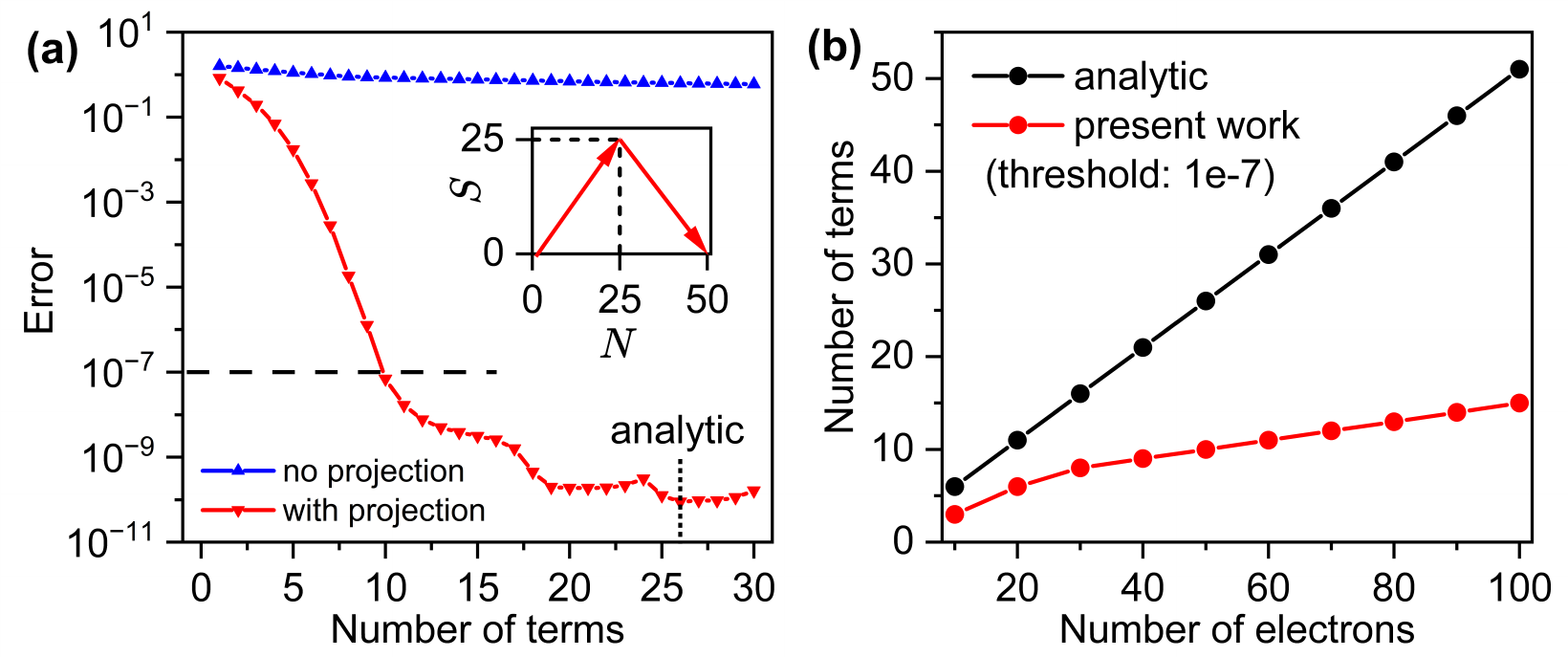}
    \caption{Tensor compression for spin eigenfunctions.
      (a) The loss function \eqref{eq: loss-CP-main} with respect to the number of terms $R$ in Eq. \eqref{eq: Theta-CP-main} (red) for the simplest singlet with $N=50$,
      in comparison with the results without the $S_z$-projection (blue).
      The inset is a schematic diagram of the singlet spin function constructed by the simplest spin coupling path.
      The horizontal black dashed line indicates the accuracy threshold of $10^{-7}$ used in this work. The vertical dotted line indicates the number of terms required by the exact analytic decomposition scheme\cite{li2025spin}.
      (b) Comparison of the number of required terms $R$ as a function of the number of electrons $N$ between the projected CP decomposition used in this work, with an accuracy threshold of $10^{-7}$, and the exact analytic decomposition scheme of Ref.~\citenum{li2025spin}.}
    \label{fig: CP}
\end{figure}

The exact sum-of-products decomposition\cite{li2025spin} of a spin eigenfunction can become expensive as $N$ grows. We therefore formulate the decomposition \eqref{eq: Theta-SAAM} as a projected CANDECOMP/PARAFAC (CP) tensor compression problem\cite{kolda2009CP}. For a target spin function $\Theta$, we approximate it by
\begin{align}
  \Theta_{\rm CP}(\sigma_1, \sigma_2, ...\sigma_N) = \sum_{r=1}^{R} C_r \prod_{j=1}^N \Big[s^r_{1j} \alpha(\sigma_j) + s^r_{2j} \beta(\sigma_j)\Big], \label{eq: Theta-CP-main}
\end{align}
where $C_r$ and $s^r_{mj}$ are fitting parameters to approximate $\Theta$ to
a sufficient accuracy. However, directly minimizing the loss $||\Theta-\Theta_{\rm CP}||^2$ 
as in standard CP decomposition does not lead to a fast decay of error with respect to the number of terms $R$, see Fig. \ref{fig: CP}a.

Since the conservation of $S_z$ can be guaranteed during the sampling process in VMC,
one only needs to fit the components where exactly $N_\alpha$ electrons are spin-up.
We can therefore minimize 
\begin{align}
  L = & || \Theta - \hat{P}_{S_z} \Theta_{\rm CP} ||^2 \nl
  = & 1 + \lan\Theta_{\rm CP}|\hat{P}_{S_z}|\Theta_{\rm CP}\ran - 2{\rm Re}\lan\Theta|\hat{P}_{S_z}|\Theta_{\rm CP}\ran, \label{eq: loss-CP-main}
\end{align} 
where $\hat{P}_{S_z}$ projects $\Theta_{\rm CP}$ to the target spin-projection sector.
For all the calculations in this work, we use the simplest spin eigenfunction constructed by the genealogical coupling scheme\cite{pauncz1979spin}, where the first $(N/2+S)$ electrons raises the total spin while the rest lowers it, see inset in Fig. \ref{fig: CP}a.
Such spin function can be represented as a matrix product state (MPS), 
which greatly reduces the cost of evaluating \eqref{eq: loss-CP-main}.
The loss \eqref{eq: loss-CP-main} is optimized using an alternating least-squares procedure\cite{kolda2009CP} described in the Supplementary Information. 

A quantitative relation between the accuracy and the number of terms $R$ is shown by the red line in Fig. \ref{fig: CP}a, taking the simplest singlet spin function with $N=50$ as an example. It is clear that with the $S_z$-projection in \eqref{eq: loss-CP-main}, the error of the CP decomposition decays significantly faster than the unprojected case, sufficiently reducing the required terms to reach a critical accuracy. In addition, compared to the exact analytic decomposition\cite{li2025spin}, where the coefficients are complex numbers, 
our approximate decomposition requires substantially less terms, see Fig. \ref{fig: CP}b, 
and uses only real numbers. This significantly reduces the computational cost 
in the evaluation of SA-NNBF wavefunction amplitudes for large systems.
All calculations in this work use compressed spin functions with the number of terms and final losses reported in Table \ref{tab: molecules}.

\subsection{Compact representation via particle-hole duality}
We note that the problem of solving $N$ electrons with $2K$ spin-orbitals is completely equivalent to the problem of solving $N_h = 2K - N$ holes\cite{li2016hilbert}. Since an electron and a hole on a same spatial site always have opposite spin, we have
$\hS_{+/-/z}^{\rm elec} = -\hS_{-/+/z}^{\rm hole}$ and hence $\hS^2_{\rm elec} = \hS^2_{\rm hole}$, implying that the $\hS^2$ operators for electrons and holes are exactly the same.
Therefore, a spin eigenfunction constructed in the hole representation is always a spin eigenfunction in the original representation with the same total spin $S$.
For active spaces that are more than half filled, the SA-NNBF amplitude can thus be evaluated using the hole occupation vector $\bn_h=(1-n_1,1-n_2,\ldots,1-n_{2K})$ whenever $N_h=2K-N<N$. This changes the determinant size from $N$ to $N_h$ and reduces the output dimension of the neural-network-generated orbital matrix. 

For iron-sulfur clusters, where the number of holes $N_h$ is much smaller than the number of electrons $N$, we find that the hole representation substantially improves the performance of the SA-NNBF ansatz compared with the electron representation. This is illustrated in Supplementary Fig.~\ref{fig: hole} for the \ce{[Fe2S2(SCH3)4]}$^{2-}$ cluster, denoted as [2Fe(III,III)-2S], using a complete active-space model with 30 electrons in 20 spatial orbitals\cite{Li2017LMO}, CAS(30e,20o). As shown in Supplementary Table~\ref{tab: hole}, for the same number of hidden units $h$, the number of variational parameters in the hole representation is less than half that in the electron representation. Nevertheless, Supplementary Fig.~\ref{fig: hole} shows that the hole representation does not lead to any loss of accuracy; instead, it gives more accurate results in most cases, especially for large $h$. In this regime, the electron representation is more prone to becoming trapped in local minima, leading to less accurate variational energies. The same trend is also observed for the original NNBF ansatz. We attribute this improvement to the reduction of redundant parameters in the hole representation for more-than-half-filled active spaces, which makes the neural-network optimization easier without significantly compromising expressivity.

We therefore use the hole representation for all more-than-half-filled active spaces in this work. For the CAS(113e,76o) active-space model of FeMoco, this particle-hole transformation reduces the determinant dimension from 113 electrons to 39 holes, as shown in Fig.~\ref{fig: scheme}(d). This reduction is essential for keeping SA-NNBF calculations accurate and efficient.

\subsection{Benchmark systems: spin contamination and scalability}

We first isolate the effect of exact spin adaptation on three benchmark systems with increasing spin complexity: a stretched H$_{12}$ chain, the active-space model of the antiferromagnetically coupled [2Fe(III,III)-2S] cluster\cite{Li2017LMO}, and a mixed-valence [2Fe(II,III)-2S] model obtained by adding one electron to the same active space. The target states are a singlet for H$_{12}$ and [2Fe(III,III)-2S], and a doublet for [2Fe(II,III)-2S]. These examples are deliberately chosen because the energy differences between spin sectors are small enough that an unconstrained variational ansatz can produce an apparently low energy while representing the wrong spin state.

\begin{figure*}[htpb]
    \centering
    \includegraphics[width=1.0\linewidth]{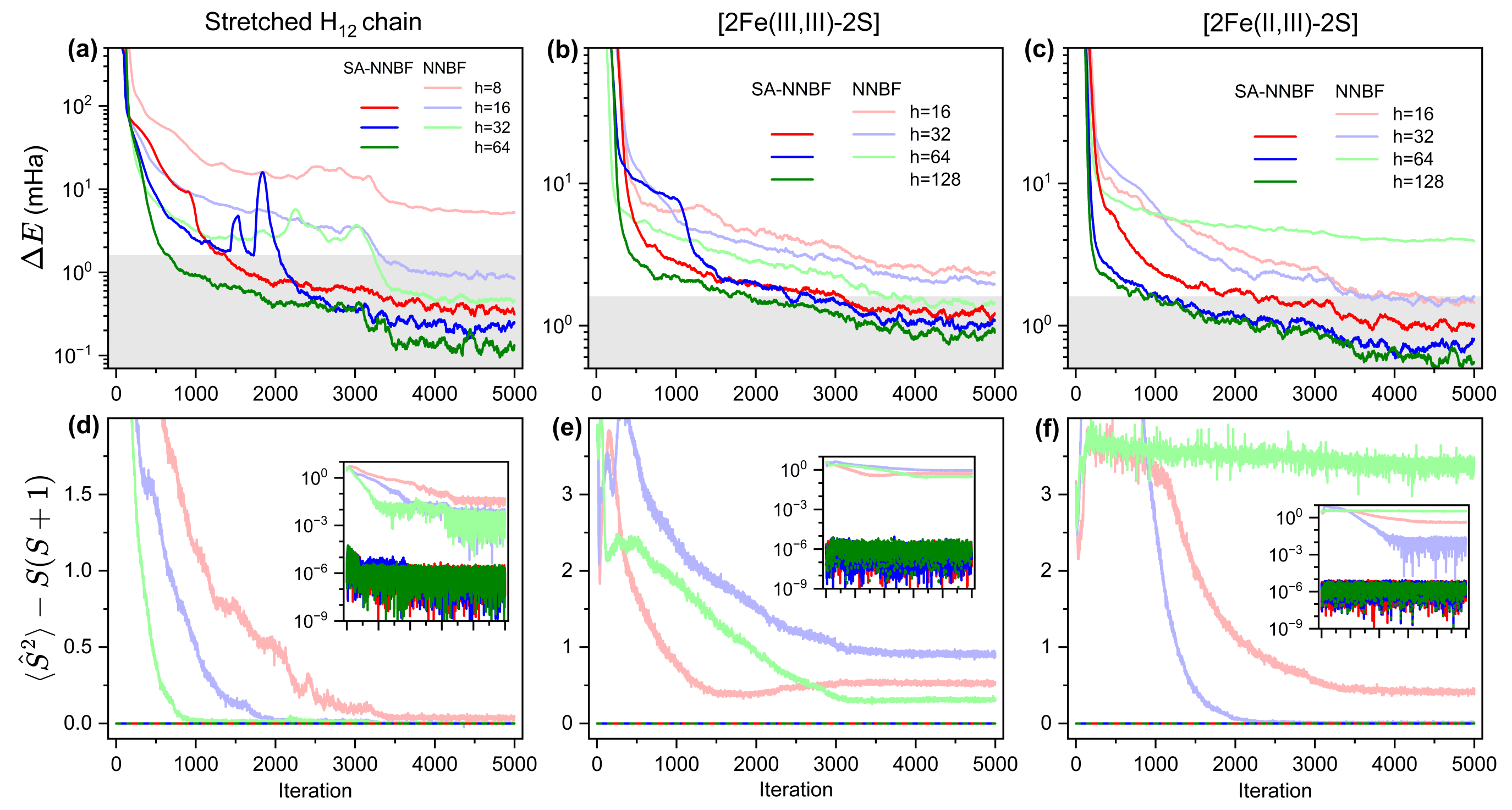}
    \caption{Energy and total spin during the VMC optimization using SA-NNBF and NNBF 
      for prototypical strongly correlated systems including stretched H$_{12}$, [2Fe(III,III)-2S] and [2Fe(II,III)-2S]. 
      The energy results are averaged over the last 100 iterations for the ease of visibility. The reference energy for \ce{H12} ($-5.66540065$ Ha) is obtained by full configuration interaction (FCI) and the ones for [2Fe(III,III)-2S] ($-116.605609$ Ha) and [2Fe(II,III)-2S] ($-116.374853$ Ha) are obtained by SA-DMRG.
      Results by SA-NNBF and NNBF are plotted with saturated and light colors, respectively.
      Curves with the same hue (red/light red, etc.) correspond to a similar number of variational parameters.
      The shaded areas in a-c represent the chemical accuracy ($<$1 kcal/mol).
      The insets in d-f show the absolute value of the total spin error in logarithmic scale.
}
    \label{fig: small}
\end{figure*}

Figure \ref{fig: small} compares SA-NNBF with standard NNBF using similar numbers of learnable parameters. Since SA-NNBF generates spatial orbitals rather than independent spin-orbitals, an SA-NNBF model with $h$ hidden units has a parameter count comparable to an NNBF model with approximately $h/2$ hidden units, as illustrated in Supplementary Table \ref{tab: hole-si}. With this parameter-matched comparison, SA-NNBF consistently gives lower variational energies than NNBF for all three systems and reaches chemical accuracy with fewer parameters.

The accompanying $\langle \hat{S}^2\rangle$ traces reveal that the energy improvement is not merely an optimization effect. SA-NNBF remains in the target spin sector throughout optimization, up to the controlled CP compression error, whereas NNBF often retains a non-negligible spin error after convergence. The effect is most pronounced in the iron-sulfur clusters, where the optimized NNBF states can contain substantial spin contamination. Thus, enforcing total-spin symmetry changes the variational landscape in a way that improves both the energy and the physical identity of the optimized state.

\begin{figure*}[htpb]
    \centering
    \includegraphics[width=1.0\linewidth]{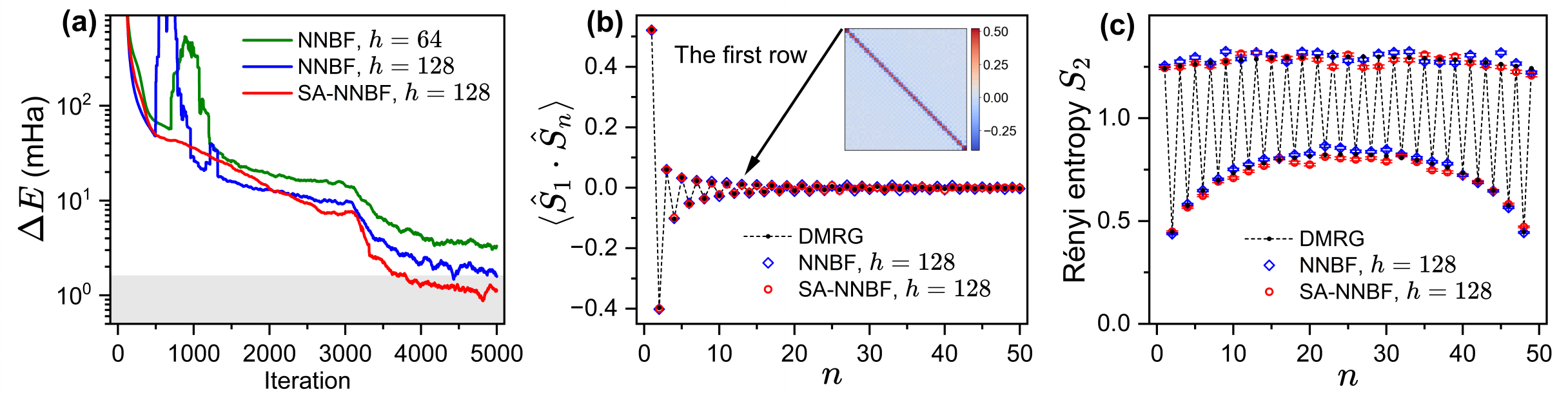}
    \caption{VMC results for the H$_{50}$ chain. 
      (a) Energy optimization curves of SA-NNBF and NNBF.
      The results are averaged over the last 100 iterations for the ease of visibility.
      The reference energy ($-26.92609$ Ha) is taken from Ref. \citenum{Hachmann2006}. 
      The shaded area shows the chemical accuracy ($<$1 kcal/mol).
      (b) Spin correlation function $\lan \hat{S}_1\cdot \hat{S}_n\ran$ between hydrogens as a function of $n$ estimated with the optimized NNBF and SA-NNBF ($h=128$), in comparison with the DMRG result as reference.
      The error bars in this subplot are smaller than the markers, and hence are not shown for clarity.
      The inset shows the diagram of the full $50\times 50$ spin correlation between different atoms calculated with the optimized SA-NNBF, where the main plot is the first row of it.
      (c) The 2-R\'enyi entropy $S_2$ as a function of the number of atoms in the first subregion starting from the leftmost hydrogen along the hydrogen chain.
}
    \label{fig: H50}
\end{figure*}

We next test whether the same symmetry-preserving construction remains useful beyond small active spaces. The H$_{50}$ chain provides a stringent one-dimensional benchmark with numerically exact DMRG data\cite{hachmann2006multireference}. As shown in Fig. \ref{fig: H50}a, SA-NNBF reaches chemical accuracy and gives a lower variational energy than a larger NNBF model.
As energy alone is not sufficient to validate a correlated wavefunction, we therefore also evaluate spin correlation functions and the 2-R\'enyi entropy of the optimized states, using the replica trick for the entropy estimator\cite{hastings2010replica}. The SA-NNBF spin correlations closely follow the DMRG reference, and the entanglement profile is reproduced across the chain. These observables show that the spin-adapted state captures not only the energy, 
but also the spatial pattern of magnetic correlations and entanglement.

\subsection{Application to FeMoco}

To further demonstrate the potential of SA-NNBF for challenging electronic-structure problems, we apply it to the active-space model of the FeMo cofactor (FeMoco) introduced by Li, Li, Dattani, Umrigar, and Chan (LLDUC)\cite{li2019electronic2}. This model contains 113 electrons in 76 localized molecular orbitals (LMOs), denoted as CAS(113e,76o). Specifically, we target the experimentally assigned ground spin state with $S=3/2$ using SA-NNBF. This system is large enough that exact diagonalization is impossible and challenging enough that recent SA-DMRG calculations require very large bond dimensions\cite{Li2025EMO}.

\begin{figure*}[t]
    \centering
    \includegraphics[width=0.8\linewidth]{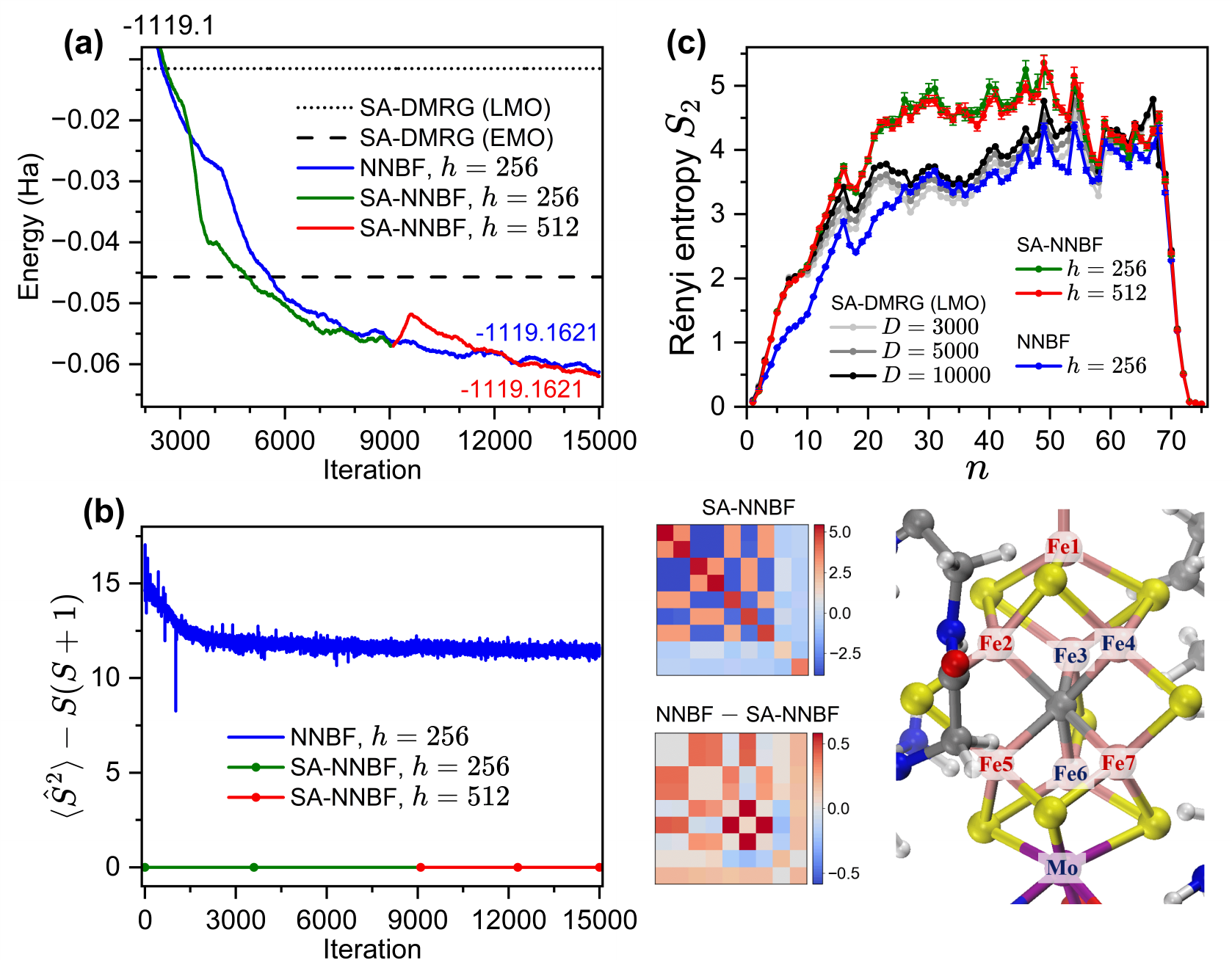}
    \caption{Energy, spin, and entanglement diagnostics for the LLDUC active-space model of FeMoco.
      (a) Energy optimization curves of SA-NNBF and NNBF in the LMO basis, 
      where the SA-NNBF optimization switches from $h=256$ to $h=512$ at the 9100-th step. The results are averaged over the last 500 iterations for ease of visibility.
      The endpoint energies are $E_{\rm NNBF}=-1119.1621\pm8.3\times 10^{-4}$ Ha and $E_{\rm SA\text{-}NNBF}=-1119.1621\pm8.6\times 10^{-4}$ Ha, estimated using an additional 819200 samples from the optimized states and rounded to the values shown in the panel.
      The SA-DMRG results with bond dimension $D=10000$ in the LMO and EMO bases\cite{Li2025EMO} are shown for comparison.       
      The core energy ($E_{\mathrm{core}}=-21021.214012$ Ha) for the LLDUC active-space Hamiltonian is not included in the reported energy.   
      Lower broken-symmetry unrestricted CC/DMRG estimates reported elsewhere\cite{zhai2026classical} are not the target of this spin-adapted comparison.
      (b) Spin-related properties. Left: Error in the total-spin expectation value. For SA-NNBF, the total spin is evaluated only at selected checkpoints.
      Middle: the group-resolved spin correlation function
      $\langle\hat{S}_A\cdot\hat{S}_B\rangle$, with the orbitals divided into nine groups: the first seven groups correspond to the seven Fe atoms, the eighth group to the Mo atom, and the last group to the remaining atoms. The upper one shows the result obtained from the optimized SA-NNBF state, while the lower one shows the difference between the NNBF and SA-NNBF results.
      Right: the structure of FeMoco, where the atoms with red labels are locally spin-up and the ones with blue are spin-down.
      The numbering of the Fe atoms follows the convention used in Ref.~\citenum{li2019electronic2}.      
      (c) The 2-R\'enyi entropy $S_2$ across bipartitions of the orbital chain, plotted as a function of the number of orbitals included in the left subsystem.
      Results for MPS wavefunctions obtained by SA-DMRG with different bond dimensions in the LMO basis are shown for comparison.}
    \label{fig: Femoco}
\end{figure*}

Figure~\ref{fig: Femoco}a shows the VMC energies. We find that both NNBF and SA-NNBF reach lower variational energies than SA-DMRG calculations with bond dimension $D=10000$, performed in both the original LMO basis and the entanglement-minimized orbital (EMO) basis\cite{Li2025EMO}. The NNBF model ($h=256$) contains 1,562,664 variational parameters, while the SA-NNBF models contain 820,382 and 1,637,790 parameters for $h=256$ and $h=512$, respectively. These parameter counts are orders of magnitude smaller than that of SA-DMRG with $D=10000$, which contains more than $10^9$ variational parameters.

Although SA-NNBF and NNBF give similar variational energies for this system, their spin properties differ substantially. As shown in Fig.~\ref{fig: Femoco}b, standard NNBF exhibits severe spin contamination: the deviation of $\langle \hat{S}^2\rangle$ from the exact value $S(S+1)$ reaches 11.3 at the end of the optimization. In contrast, SA-NNBF preserves the target total spin by construction and yields the correct value of $\langle \hat{S}^2\rangle$ throughout the calculation. Furthermore, the inset of Fig.~\ref{fig: Femoco}b shows that NNBF systematically overestimates the inter-group spin correlation functions $\langle \hat{S}_A\cdot\hat{S}_B\rangle$, particularly for negatively correlated pairs. This behavior is consistent with its erroneous total-spin expectation value, since the total spin satisfies $\langle \hat{S}^2\rangle=\sum_{AB}\langle \hat{S}_A\cdot\hat{S}_B\rangle$. These results show that, for FeMoco, variational energy alone is insufficient to assess wavefunction quality, and that  spin adaptation is essential for obtaining physically meaningful spin correlations.

To understand the energetic advantage of SA-NNBF over SA-DMRG, we further compute the 2-R\'{e}nyi entropies $S_2$ across bipartitions along the MPS chain, which provide a measure of the entanglement encoded in the wavefunction. As shown in Fig.~\ref{fig: Femoco}c, the SA-DMRG wavefunctions exhibit increasing $S_2$ as the bond dimension $D$ is enlarged as expected. 
The optimized SA-NNBF state yields larger $S_2$ values than SA-DMRG near the middle of the MPS chain, while showing good agreement with SA-DMRG near the two boundaries. 
This suggests that SA-NNBF can represent a more entangled state than the currently accessible SA-DMRG calculations in the most entangled region of the orbital chain. 
Its ability to encode such entanglement compactly may therefore underlie the lower variational energy achieved by SA-NNBF. 
By contrast, the NNBF state does not show larger $S_2$ values than SA-DMRG near the middle of the chain and fails to reproduce the SA-DMRG entanglement profile near the left boundary. 
This qualitative discrepancy, despite its low variational energy, further indicates that the standard NNBF wavefunction suffers from an unphysical description of the FeMoco spin state.

Finally, we note that our FeMoco comparison with DMRG is made within the spin-adapted variational framework. Although the present SA-NNBF energy remains above the latest lower-energy estimate obtained by combining unrestricted coupled cluster and unrestricted DMRG (CC/DMRG) in a broken-symmetry orbital framework\cite{zhai2026classical}, SA-NNBF achieves a lower energy than recent SA-DMRG calculations at $D=10000$ while preserving the target total spin exactly. This highlights the ability of a compact neural-network ansatz to capture substantial correlation and entanglement in FeMoco under an exact spin constraint. 
More expressive architectures, such as Transformers\cite{Viteritti2023transformer,Shang2025transformer,gu2025solving,rende2026transformer}, or multiple backflow orbital sets may further improve the variational energy while retaining spin purity.

\section{Discussion}

We have introduced a second-quantized neural-network wavefunction that enforces total-spin symmetry at the ansatz level while retaining the flexibility of neural-network backflow. The central message is that spin symmetry is not only a formal property for strongly correlated molecules. In the examples studied here, especially the FeMoco, a flexible but spin-unconstrained ansatz can obtain a low variational energy while representing the wrong spin state and distorted spin correlations. SA-NNBF removes this ambiguity by ensuring that the optimized state remains in the target spin sector throughout the VMC calculation, with the residual error set by a controlled spin-function compression.
The projected tensor compression algorithm and particle-hole representation introduced here allows to reach active spaces with more than one hundred electrons. The FeMoco calculation demonstrates the resulting capability: a compact spin-adapted neural-network state can be competitive with large-bond-dimension spin-adapted DMRG.

Several directions remain open. The present implementation intentionally uses a simple one-hidden-layer feed-forward network to isolate the effect of spin adaptation. More expressive architectures, including Transformer-based backflow models\cite{Viteritti2023transformer,Shang2025transformer,gu2025solving,rende2026transformer}, multiple sets of spatial orbitals, and richer spin-coupling paths could further lower the variational energy. 
Looking forward, the same spin-adapted framework can be extended to excited-state VMC by targeting multiple spin sectors or enforcing orthogonality between states\cite{Choo2018Symmetries,Entwistle2023excited,Pfau2024excited}. 
Besides, the second-quantized formulation further makes SA-NNBF naturally compatible with quantum embedding frameworks\cite{Sun2016Embedding}, in which large molecular or materials systems are mapped onto effective correlated active spaces embedded in an environment. This opens a route to applying spin-adapted neural-network quantum states to much larger systems, while retaining an explicit and exact treatment of total-spin symmetry in the embedded strongly correlated region. 

\section{Methods}
\subsection{Variational Monte Carlo optimization}
We use the variational Monte Carlo\cite{becca2017quantum} (VMC) method to solve the electronic Schr\"odinger equation in second quantization
\begin{equation}
    \hat{H}|\Psi\rangle = E|\Psi\rangle,
\end{equation}
where $\hat{H}$ is the ab initio molecular Hamiltonian
\begin{equation}
    \hat{H} = \sum_{pq}h_{pq}\hat{a}_p^{\dagger}\hat{a}_q + \frac{1}{4} \sum_{pqrs}\langle pq \| rs \rangle
    \hat{a}_{p}^{\dagger}\hat{a}_q^{\dagger} \hat{a}_s\hat{a}_r,\label{eq:Ham}
\end{equation}
with $h_{pq}$ and $\langle pq\|rs\rangle$ being one-electron
and two-electron molecular integrals, respectively,
$\hat{a}_p^{(\dagger)}$ being the fermionic annihilation (creation) operator for the $p$-th
spin-orbital.
We represent $|\Psi\rangle$ with an NQS ansatz,
\begin{equation}
    |\Psi_\theta\rangle = \sum_{\bm{n}} \Psi_\theta(\bm{n}) |\bm{n}\rangle,
\end{equation}
where $\theta$ denotes the set of variational parameters.
In VMC, the variational energy can be estimated as
\begin{equation}
    E_\theta =
    \frac{\langle\Psi_{\theta}|\hat{H}|\Psi_{\theta}\rangle}{\langle\Psi_{\theta}|\Psi_{\theta}\rangle}
    = \langle E_{\rm{loc}}(\bm{n})\rangle_{\bm{n}\sim P_\theta(\bm{n})},
\end{equation}
where ${P_{\theta}(\bm{n})}$ is the probability distribution $P_{\theta}(\bm{n}) = |\Psi_{\theta}(\bm{n})|^2/\sum_{\bm{n}} |\Psi_{\theta}(\bm{n})|^2$ and $E_{\rm{loc}}(\bm{n})$ is the local energy defined by
\begin{equation}
    E_{\rm{loc}}(\bm{n}) =  \frac{\langle \bm{n}|\hat{H}|\Psi_\theta\rangle}{\langle \bm{n}|\Psi_\theta\rangle} =
    \sum_{\bm{m}} H_{\bm{n}\bm{m}}\frac{\Psi_\theta(\bm{m})}{\Psi_\theta(\bm{n})}.\label{eq:exactElocal}
\end{equation}
Here, $H_{\bm{n}\bm{m}}$ is the matrix representation of Hamiltonian \eqref{eq:Ham} in the occupation-number representation. Following from Eq. \eqref{eq:Ham}, $H_{\bm{n}\bm{m}}$ is sparse,
with $O(K^4)$ nonzero elements per row. Sampling $\bm{n}$ according to ${P_{\theta}(\bm{n})}$ can be obtained using Markov chain Monte Carlo (MCMC)\cite{Levin2017_Markov}.
The energy gradient with respect to parameters $\theta$ can be evaluated by\cite{becca2017quantum}
\begin{equation}
    \partial_{\theta} E_{\theta}  = 2\Re\big[ \big\langle(\partial_\theta\ln{\Psi_\theta^*(\bm{n})}) (E_{\rm {loc}}(\bm{n}) - E_\theta) \big\rangle_{\bm{n}\sim P_\theta(\bm{n})}\big],\label{eq:egrad}
\end{equation}
where $\partial_\theta\ln{\Psi_\theta^*(\bm{n})}$ can be calculated using automatic differentiation (AD) techniques\cite{Griewank2008_AD}. The parameters $\theta$ are updated according to $\partial_{\theta} E_{\theta} $ using an appropriate optimizer, such as 
AdamW\cite{loshchilov2019fixing}, stochastic reconfiguration (SR)\cite{sorella1998green}, or more advanced ones\cite{chen2024empowering,rende2024simple,goldshlager2024kaczmarz,gu2025solving}.

In all production calculations, the Markov-chain moves are restricted to the target particle-number and $S_z$ sector. Total-spin adaptation is supplied by the spin function in SA-NNBF, with the residual symmetry error controlled by the projected CP compression loss reported in Supplementary Table \ref{tab: molecules}. The optimizer, sampling exponent, local-energy threshold, number of stochastic local-energy samples, burn-in length and FeMoco pretraining protocol are summarized in Supplementary Tables \ref{tab: molecules} and \ref{tab: opt}.

\subsection{Neural-network architecture for spatial backflow orbitals}\label{subsec: NN}
The neural network architecture for generating spatial orbitals $\bUbar(\bnbar)$ used in this work is constructed as follows. For each occupation number vector $\bn$, we first calculate the corresponding spatial occupation number vector $\bnbar$, and encode each element $\nbar_j = 0, 1, 2$ into a length-3 vector with a learnable $3 \times 3$ matrix (embedding layer), resulting in a vector $\bm{m}$ with length $3K$.
Then, $\bm{m}$ is passed to a 1-layer FNN
\begin{align}
  \bm{x}_h = {\rm SiLU}(\mathbf{W}_1 \bm{m} + \bm{b}_1),
\end{align}
where $\mathbf{W}_1$ and $\bm{b}_1$ are the weight and bias, respectively, $\bm{x}_h$ is the state of the hidden layer with length $h$, and SiLU is the sigmoid linear unit function.
Finally, $\bm{x}_h$ is transformed to the output layer as
\begin{align}
  \bm{u} = \mathbf{W}_2 \bm{x}_h + \bm{b}_2,
\end{align}
which is reshaped into the $K \times N$ spatial orbital coefficient matrix $\bUbar$.

The same architecture is used for all systems so that the effect of spin adaptation can be isolated from architectural changes. The hidden dimension $h$ and numerical precision are listed in Supplementary Tables \ref{tab: molecules} and \ref{tab: opt}. For calculations in the hole representation, the input occupation vector and the output orbital matrix are replaced by their hole-space counterparts, as described above.

\subsection{Semi-stochastic local-energy evaluation}

For short-range model Hamiltonians, local-energy evaluation often scales linearly with the number of sites. In ab initio molecular Hamiltonians, by contrast, each occupation vector can be connected to $O(K^4)$ single and double excitations. Exact evaluation of Eq. \eqref{eq:exactElocal} therefore requires $O(N_{\rm{s}}K^4)$ Hamiltonian-element evaluations and, more importantly, the same order of NQS amplitude evaluations $\Psi_\theta(\bm{m})$, where $N_{\rm{s}}$ is the number of unique samples. This amplitude-evaluation cost is the main bottleneck for applying NQS-based VMC to large molecular active spaces.

In this work, we use the semi-stochastic algorithm \cite{wu2025hybrid} proposed in our previous work to reduce the computational cost.
The main idea is to decompose the local energy into two parts based on the magnitude of 
$H_{\bm{n}\bm{m}}$. Given a threshold $\epsilon$, which is a hyperparameter
in this scheme, the deterministic part $E_{\rm{loc}}^{\rm{d}}(\bm{n}, \epsilon)$ involves summing over all the matrix elements $H_{\bm{n}\bm{m}}$ that satisfy $|H_{\bm{n}\bm{m}}| \geq \epsilon$, viz.,
\begin{equation}
    E_{\rm{loc}}^{\rm{d}}(\bm{n}, \epsilon) = \sum_{\{ \bm{m} \,:\, |H_{\bm{n}\bm{m}}| \geq \epsilon\}} H_{\bm{n}\bm{m}}\frac{\Psi_\theta(\bm{m})}{\Psi_\theta(\bm{n})}. 
\end{equation}
The stochastic part $E_{\rm{loc}}^{\rm{s}}(\bm{n}, \epsilon, N_{\epsilon})$ is designed to handle the smaller contributions by sampling $\bm{m}^{\prime}$ from the distribution $ P_{\bm{n}}(\bm{m}^{\prime}) \propto |H_{\bm{n}\bm{m}^{\prime}}|$, where $\bm{m}'$ is defined by $|H_{\bm{n}\bm{m}^{\prime}}| < \epsilon$. Specifically, the evaluation of the stochastic part can be expressed as
\begin{equation}
    E_{\rm {loc}}^{\rm{s}}(\bm{n}, \epsilon, N_{\epsilon}) = 
    \Big\langle 
    \frac{H_{\bm{n}\bm{m}^{\prime}}}{P_{\bm{n}}(\bm{m}^{\prime})}\frac{\Psi_\theta(\bm{m}^{\prime})}{\Psi_\theta{(\bm{n})}}
    \Big\rangle_{\bm{m}^\prime \sim P_{\bm{n}}(\bm{m}')},
\end{equation}
where $N_{\epsilon}$ is the number of samples that can be chosen based on the desired accuracy. 

The final local energy $E_{\rm{loc}}(\bm{n})$ is evaluated as
\begin{equation}
    E_{\rm{loc}}(\bm{n}, \epsilon, N_{\epsilon}) = E_{\rm{loc}}^{\rm{d}}(\bm{n}, \epsilon) + E_{\rm loc}^{\rm{s}}(\bm{n}, \epsilon, N_{\epsilon}), 
\end{equation}
which is an unbiased estimator of the energy. This decomposition significantly reduces the number of wavefunction amplitudes to be calculated, which would otherwise be computationally expensive. 
In practice, it is necessary to choose $\epsilon$ and $N_{\epsilon}$ in order to strike a balance between the variance and the computational complexity. 
The accuracy and cost tradeoff for the chosen $\epsilon$ and $N_{\epsilon}$ values is benchmarked in Supplementary Fig. \ref{fig: eloc}. For the systems studied here, the semi-stochastic estimator reduces local-energy evaluation cost by one to three orders of magnitude while keeping the mean local-energy bias below the stochastic resolution used in the VMC optimizations.

\section*{Data Availability}
The data supporting the findings of this study are available within the article and Supplementary Information. Molecular-integral files for the iron-sulfur clusters are publicly available on Github\cite{linkToFCIDUMPfe2fe4,linkToFCIDUMPfemoco}.

\section*{Code Availability}

The implementation is available in the \textsc{PyNQS} code base at https://github.com/Quantum-Chemistry-Group-BNU/PyNQS. Example input files and scripts for the SA-NNBF calculations are available at https://github.com/Lex-Y-Darkblade/SA-NNBF\_examples.

\section*{Acknowledgment}
The authors acknowledge helpful discussion with Ruichen Li, Weiluo Ren, and Dingshun Lv. This work was supported by the Quantum Science and Technology-National Science and Technology Major Project (2023ZD0300200) and the Fundamental Research Funds for the Central Universities.

\section*{Author contributions}
Y.L. and Z.L. conceived the research. Y.L. wrote the code, performed the experiments, and wrote the paper. Z.W. and B.Z. assisted in writing the code and preparing the manuscript. W.F. and Z.L. oversaw the entire project. All authors contributed to the discussion of the results.

\section*{Competing interests}
The authors declare no competing interests.

\bibliographystyle{apsrev4-2}
\bibliography{main}

\let\addcontentsline\oldaddcontentsline

\pagebreak
\clearpage
\raggedbottom
\pagebreak

\widetext
\allowdisplaybreaks[4]
\begin{center}
\textbf{\large Supplementary Information for \\
``Spin-adapted neural network backflow for strongly correlated electrons"}\\
\vspace{2ex}
Yunzhi Li$^{1,2}$, Zibo Wu$^{1,2}$, Bohan Zhang$^{1,2}$, Wei-Hai Fang$^{1,2}$, and Zhendong Li$^{1,2,*}$ \\
{\it $^1$ Key Laboratory of Theoretical and Computational Photochemistry, Ministry of Education, College of Chemistry, Beijing Normal University, Beijing, 100875, China \\
$^2$ Institute for Advanced Study, Beijing Normal University, Beijing, 100875, China}
\end{center}

\setcounter{secnumdepth}{3}
\setcounter{section}{0}
\setcounter{equation}{0}
\setcounter{figure}{0}
\setcounter{table}{0}
\setcounter{page}{1}
\makeatletter
\renewcommand{\thesection}{S\arabic{section}}
\renewcommand{\thefigure}{S\arabic{figure}}
\renewcommand{\thetable}{S\arabic{table}}
\renewcommand{\appendixname}{}
\counterwithout{equation}{section} 
\renewcommand{\theequation}{S\arabic{equation}}

\tableofcontents

\subsection{Details of the projected CP decomposition for spin eigenfunctions}
An arbitrary normalized $N$-electron spin eigenfunction of both $\hat{S}^2$ and $\hat{S}_z$ can be encoded as a rank-$N$ tensor as
\begin{align}
  \Theta(\sigma_1, \sigma_2, ...\sigma_N) = \sum_{m_1, m_2, ..., m_N} A_{m_1, m_2, ..., m_N} \prod_{i=1}^{N} \gamma_{m_i}(\sigma_i), \label{eq: Theta}
\end{align} 
where $m_i = 0, 1$, $\gamma_0(\sigma)\equiv\alpha(\sigma)$, and $\gamma_1(\sigma)\equiv\beta(\sigma)$.
We attempt to express $\Theta$ by a sum-of-product form $\Theta_{\rm CP}$
\begin{align}
  \Theta_{\rm CP}(\sigma_1, \sigma_2, ...\sigma_N) = 
  \sum_{m_1, m_2, ..., m_N} \sum_{r=1}^{R} C_r \prod_{j=1}^N s^r_{m_j j} \gamma_{m_j}(\sigma_j). \label{eq: Theta-CP}
\end{align}
The projection operator $\hat{P}_{S_z}$ can be represented as
\begin{align}
  \hat{P}_{S_z} = \delta_{(\sum_j m_j),(N/2 - S_z)} \equiv \delta_{MN_\beta}, \label{eq: PSz}
\end{align}
where $M \equiv \sum_{j} m_j$ denotes the number of spin-down electrons, $N_{\beta} \equiv N/2 - S_z$ is the target number of spin-down electrons for the specified $S_z$ value. 
Substituting Eqs. \eqref{eq: Theta}, \eqref{eq: Theta-CP} and \eqref{eq: PSz} into \eqref{eq: loss-CP-main}, one can evaluate the last two terms in \eqref{eq: loss-CP-main} as
\begin{align}
  \lan\Theta_{\rm CP}|\hat{P}_{S_z}|\Theta_{\rm CP}\ran
  = & \sum_{m_1, m_2, ..., m_N} \delta_{MN_\beta} \sum_{r=1}^{R} C_r \sum_{t=1}^{R} C_t \prod_{j=1}^N s^r_{m_j j} s^{t}_{m_j j} \nl
  = & \sum_{m_1, m_2, ..., m_N} \frac{1}{N+1} \sum_{k=0}^{N} e^{\ii 2\pi k (\sum_i m_i - N_\beta)/(N+1)} \sum_{r=1}^{R} C_r \sum_{t=1}^{R} C_t \prod_{j=1}^N s^r_{m_j j} s^{t}_{m_j j} \nl 
  = 
  &\frac{1}{N+1}\sum_{k=0}^{N} e^{- \ii 2\pi k N_\beta/(N+1)} \sum_{r,t=1}^{R} C_r C_t \prod_{j=1}^N \Big[ \sum_{m_j=0}^1  s^r_{m_j j} s^{t}_{m_j j} e^{\ii 2\pi k m_j/(N+1)} \Big], \label{eq: CP-P-CP}
\end{align}
and
\begin{align}
  \lan\Theta|\hat{P}_{S_z}|\Theta_{\rm CP}\ran
  = & \sum_{m_1, m_2, ..., m_N} A_{m_1, m_2, ..., m_N}^{*} \delta_{MN_\beta} \sum_{r=1}^{R} C_r \prod_{j=1}^N s^r_{m_j j} \nl
  = & \sum_{m_1, m_2, ..., m_N} \frac{1}{N+1} \sum_{k=0}^{N} e^{\ii 2\pi k (\sum_i m_i - N_\beta)/(N+1)} A_{m_1, m_2, ..., m_N}^{*} \sum_{r=1}^{R} C_r \prod_{j=1}^N s^r_{m_j j} , \label{eq: Theta-P-CP}
\end{align}
where we have used the Fourier expansion of Kronecker delta
\begin{align}
  \delta_{nn'} = \frac{1}{N+1} \sum_{k=0}^{N} e^{\ii 2\pi k (n-n')/(N+1)},
\end{align}
for $n, n' = 0, 1, ..., N$. If the target state $\Theta$ can be efficiently written as a matrix product state
\begin{align*}
  A_{m_1, m_2, ..., m_N} = \sum_{\alpha_1, \alpha_2,...,\alpha_{N}} A^{m_1}_{\alpha_1} A^{m_2}_{\alpha_2\alpha_3} A^{m_3}_{\alpha_3\alpha_4} ... A^{m_{N-1}}_{\alpha_{N-1}\alpha_{N}} A^{m_{N}}_{\alpha_{N}},
\end{align*}
which is exactly the case when $\Theta$ in \eqref{eq: Theta} is constructed by the genealogical coupling scheme\cite{pauncz1979spin}, then the evaluation of Eqs. \eqref{eq: CP-P-CP} and \eqref{eq: Theta-P-CP} can be achieved polynomially in $N$.


Substituting Eqs. \eqref{eq: CP-P-CP} and \eqref{eq: Theta-P-CP} into \eqref{eq: loss-CP-main}, we notice that for fixed values of all CP factors except those on site $j$, the cost function $L$ is \emph{quadratic} in the parameter set of every specific $j$, viz., $\{s^r_{m_{j}j}|r=1,2,...,R; m_j=0,1\}$.
Thus, the minimization problem reduces to linear least-squares problem.
Following the idea of alternating least square\cite{kolda2009CP} (ALS) in CP decomposition, one can minimize the cost by iteratively by sweeping over all the sites $j$ to obtain an optimal sum-of-product fitting of $\Theta$ for a given number of terms $R$. The reported value of $L$ is the final projected error in the sampled $S_z$ sector,
see Eq. \eqref{eq: loss-CP-main}; in the limit $L\rightarrow0$, the compressed spin function recovers the target spin eigenfunction in that sector exactly.

%
%

\subsection{Particle-hole representation benchmark}

\begin{figure}[htpb]
    \centering
    \includegraphics[width=0.45\linewidth]{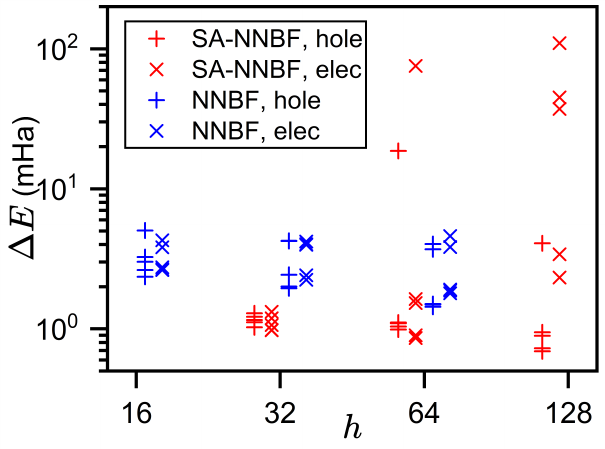}
    \caption{Errors of SA-NNBF and NNBF for the CAS(30e,20o) model\cite{Li2017LMO} of the [2Fe(III,III)-2S] cluster in the hole and electron representations for different hidden-layer sizes $h$. For each $h$, five independent calculations were performed with a given ansatz.}
    \label{fig: hole}
\end{figure}

This section provides the detailed benchmark supporting the use of the particle-hole representation in the main text, see Table \ref{tab: hole-si} and Fig. \ref{fig: hole}. The comparison is performed for the [2Fe(III,III)-2S] CAS(30e,20o) active-space model\cite{Li2017LMO}, where the number of holes is substantially smaller than the number of electrons.

\begin{table}[htpb]
\centering
\caption{Numbers of variational parameters in SA-NNBF and NNBF in hole and electron representations for the [2Fe(III,III)-2S] cluster.}
\label{tab: hole}
\label{tab: hole-si}
\begin{tabular}{ccccc}
  \hline\hline
  \multirow{2}{*}{$h$} & \multicolumn{2}{c}{SA-NNBF} & \multicolumn{2}{c}{NNBF} \\
  & hole & electron & hole & electron \\
  \midrule
  16 &  &  & 7456 & 21056 \\
  32 & 8562 & 21762 & 14512  & 40912 \\
  64 & 16914 & 42914 & 28624 & 80624 \\
  128 & 33618 & 85218 &  & \\
  \hline\hline
\end{tabular}
\end{table}

\subsection{Semi-stochastic local-energy benchmark}

\begin{figure*}[htpb]
    \centering
    \includegraphics[width=0.9\linewidth]{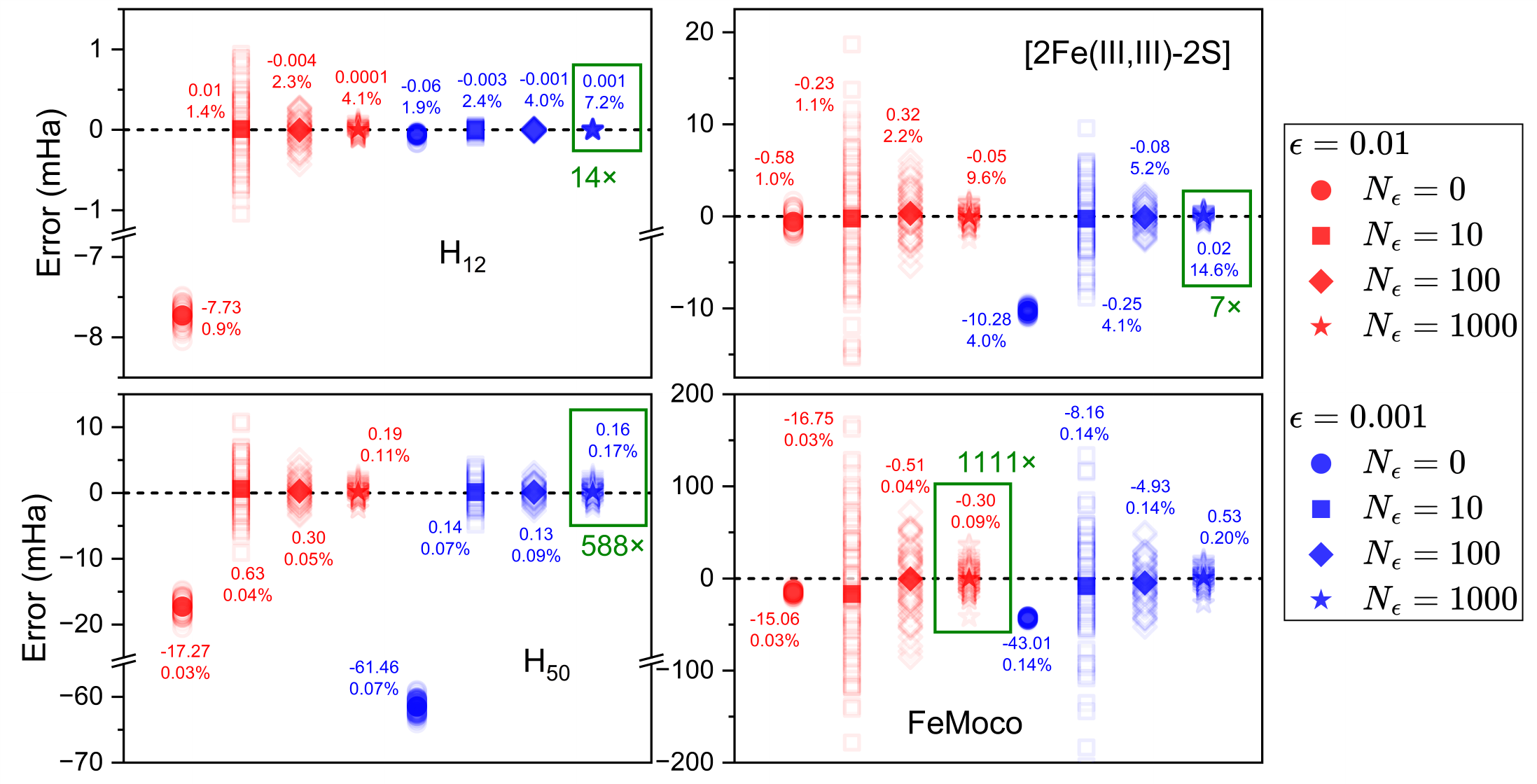}
    \caption{Benchmark for different hyperparameters $\epsilon$ and $N_{\epsilon}$ in the semi-stochastic local-energy evaluation for the systems considered in this work.
      For each combination of $\epsilon$ and $N_{\epsilon}$, 100 batches of samples are generated, with 4096 samples per batch for \ce{H12}, [2Fe(III,III)-2S], and \ce{H50}, and 8192 samples per batch for FeMoco.
      The plotted error is the difference between the resulting mean local energy of each batch and the exact mean local energy of the same batch.
      Labels show the mean error over all batches in milli-Hartree and the computational cost relative to exact local-energy evaluation.
      Green boxes mark the hyperparameters used in the production calculations, with the corresponding speedups shown in green.
}
    \label{fig: eloc}
\end{figure*}

We benchmark the semi-stochastic local-energy estimator by comparing batch-averaged local energies against exact local-energy evaluation for the same batches. This test determines the $\epsilon$ and $N_{\epsilon}$ values used in the production calculations, see Fig. \ref{fig: eloc}.

\subsection{Implementation and computational details}
We implemented the SA-NNBF ansatz in the \textsc{PyNQS}\cite{wu2025hybrid,pynqs_github} package based on PyTorch\cite{paszke2019pytorch}. Computational details for the molecular systems considered in this work are summarized in Table \ref{tab: molecules}.
For hydrogen chains, the orthonormalized atomic orbitals (OAOs) are used as the one-electron orbitals\cite{Hachmann2006}, while for iron-sulfur clusters, we use the active space models with localized molecular orbitals (LMOs) constructed in previous works\cite{Li2017LMO, li2019electronic2}, whose molecular integrals are publicly available on GitHub\cite{linkToFCIDUMPfe2fe4,linkToFCIDUMPfemoco}.

\begin{table}[htpb]
\renewcommand{\arraystretch}{1.0}
\centering
\caption{Computational details for molecules investigated in this work.}
\label{tab: molecules}
\begin{tabular}{ccccc}
  \hline\hline 
  & H$_{12}$ chain & [2Fe(III,III)-2S], [2Fe(II,III)-2S] & H$_{50}$ chain & FeMoco \\
  \midrule
  Geometry (bond length) & 4.0 Bohr & Ref. \citenum{Li2017LMO, li2019electronic2} & 2.0 Bohr & Ref. \citenum{Li2017LMO, li2019electronic2} \\
  Basis & STO-3G, OAO & LMO \cite{Li2017LMO, li2019electronic2} & STO-6G, OAO & LMO \cite{Li2017LMO, li2019electronic2} \\
  No. of active spatial orbitals & 12 & 20 & 50 & 76 \\
  No. of active electrons & 12 & 30, 31 & 50 & 113 \\
  Precision & float64 & float64 & float32 & float32 \\
  MCMC walkers & 4096 & 4096 & 4096 & 8192 \\
  MCMC starting state & random & random & random & MPS \\
  MCMC burn-in & 2500 & 2500 & 2500 & 3000 \\
  Sampling exponent $\alpha$ & 2.0 & 2.0 & 1.5 & 2.0 \\
  Local energy threshold $\epsilon$ & 0.001 & 0.001 & 0.001 & 0.01 \\
  Local energy samples $N_{\epsilon}$ & 1000 & 1000 & 1000 & 1000 \\
  Terms in exact decomposition\cite{li2025spin} & 7 & 6, 5 & 26 & 19 \\
  Terms used in $\Theta_{\rm CP}$ & 4 & 3, 3 & 10 & 12 \\
  Error $L$ of $\Theta_{\rm CP}$ & $2.6\times10^{-10}$ & $1.0\times10^{-10}$, $9.0\times10^{-11}$ & $7.2\times10^{-8}$ & $2.7\times10^{-8}$ \\
  \hline\hline
\end{tabular}
\end{table}

\begin{figure}[htpb]
    \centering
    \includegraphics[width=0.8\linewidth]{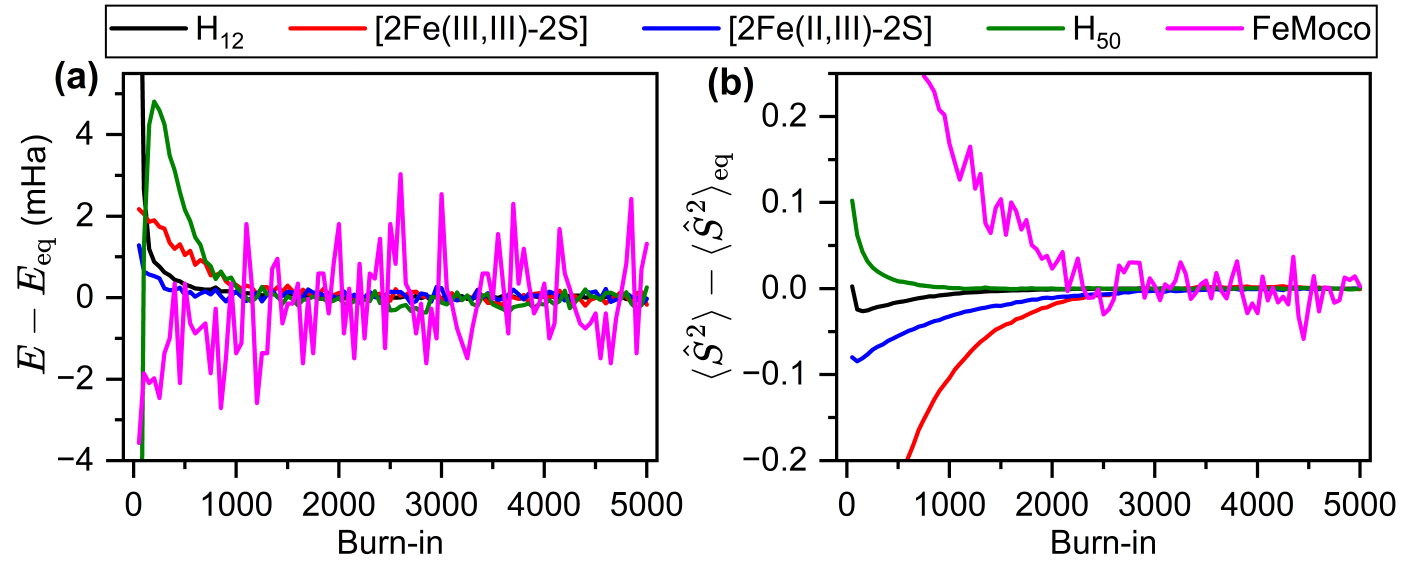}
    \caption{Benchmark results for the estimations of (a) energy and (b) total spin with respect to the MCMC burn-in steps. 
      They are estimated by running 409600 MCMC trajectories sampled from optimized NNBF states. The equilibrium (eq) values are the averages over steps 4550, 4600, $\cdots$, 5000 (in total 409600$\times$10 samples).
}
    \label{fig: burn}
\end{figure}

In the MCMC sampling, we propose trial moves on each walker by applying a single excitation on the current configuration. For each MCMC chain, we take one sample per walker after sufficient number of burn-in steps from an initial configuration. 
Benchmark results for the MCMC burn-in steps are displayed in Fig. \ref{fig: burn}.
In this work, the MCMC initial configurations for systems except FeMoco are generated as a uniformly random state, while the ones for FeMoco are generated by sampling from an auxiliary matrix product state (MPS) obtained with a bond dimension of 100.

For most of the molecules, configurations are sampled according to the absolute squares of the wavefunction values, $|\Psi(\bn)|^2$.
However, it is suggested\cite{Misery2026} that sampling according to a distribution of $|\Psi(\bn)|^\alpha$ and then reweighting the samples by a factor of $|\Psi(\bn)|^{2-\alpha}$ can sometimes improve performance, where the optimal $\alpha$ is usually less than 2. In this work, this technique is used in calculations on the H$_{50}$ chain, where we adjust $\alpha$ to 1.5. 

The number of terms in the decomposed spin wavefunction $\Theta_{\rm CP}$ for each molecule is also listed in Table \ref{tab: molecules}, where the number of terms obtained with the exact decomposition\cite{li2025spin} is also shown for comparison. 

For the optimization of the NQS models, we use a combination of MinSR\cite{chen2024empowering,Rende2024} and AdamW\cite{loshchilov2019fixing}, where we pass the MinSR output to the AdamW optimizer to evaluate the final updates on the parameters. 
For molecules except FeMoco, each NQS is optimized by 5000 steps from a random initial state, with a learning rate which is constant in the first 3000 steps and exponentially decays in the last 2000 steps, see Table \ref{tab: opt} for details.
For FeMoco, the NQSs are pretrained by an MPS with a bond dimension of 100 before the optimization. The SA-NNBF for FeMoco is optimized through four stages with each containing $3000\sim 6000$ steps and summing up to 15000 steps, 
where the learning rate schedule in each stage is similar to the one for other molecules except for an additional linear warm-up at the beginning,
and the neural network size is expanded from $h=256$ to $h=512$ at the beginning of the third stage (9100-th step). The NNBF for FeMoco is optimized through a single stage with the following learning rate schedule
\begin{align*}
  \begin{cases}
    10^{-3}\cdot t/2000, & t \le 2000, \\
    10^{-3}, & t \le 4000, \\
    10^{-3}/[1+(t-4000)/1000], & t \le 13000, \\
    10^{-4} / 10^{(t-13000)/2000}, & t \le 15000,
  \end{cases}
\end{align*}
where $t$ denotes the current step.

\begin{table}[htpb]
\renewcommand{\arraystretch}{1.0}
\centering
\caption{Computational details for the optimization. For FeMoco, the complete optimization consists of several stages, each using a learning-rate schedule similar to that used for the other molecules.}
\label{tab: opt}
\begin{tabular}{cc}
  \hline\hline 
  Hyperparameter & Choice \\
  \midrule
  Optimizer & MinSR + AdamW \\
  MinSR damping $\lambda$ & $10^{-3}$ \\
  AdamW $\beta_1$ & 0.9 \\
  AdamW $\beta_2$ & 0.999 \\
  optimization steps & 5000 (except FeMoco); 15000 (FeMoco)\\
  learning rate & $\min[10^{-3}, 10^{-3} / 10^{(t-3000)/1000}]$ (except FeMoco)\\
  \hline\hline 
\end{tabular}
\end{table}

%
%

\end{document}